\documentclass[prd,aps,12pt]{revtex4}
\usepackage[dvips]{graphicx}
\usepackage{slashed}
\usepackage{amssymb}
\usepackage{natbib}
\usepackage{amsmath}
\usepackage{url}
\usepackage{color}
\begin{document}
\renewcommand{\baselinestretch}{1.5}
\newcommand{\Prd}{Phys. Rev D}
\newcommand{\Prl}{Phys. Rev. Lett.}
\newcommand{\Pl}{Phys. Lett.}
\newcommand{\Cqg}{Class. Quantum Grav.}
\newcommand{\Sch}{Schwarzschild$\;$}
\def \vx{\vec x}
\def \v{{\vec v}}
\def \vomega{\vec \omega}
\def \vnabla{\vec\nabla}
\def \vxi{\vec\xi}
\def\d{{\mathrm{d}}}
\def\half{\frac 1 2}
\def\a{\alpha}
\def\b{\beta}
\def\part{\partial}

\def\setR{\mathbb{R}}
\def\setN{\mathbb{N}}
\def\setC{\mathbb{C}}
\def\calH {{\cal H}}
\def\ie {{i.e.}}
\def\sgn{\mathrm{sgn}}

\def \hz{{\hat {\bf z}}}
\def \hx{{\hat {\bf x}}}
\def \hy{{\hat {\bf y}}}
\newcommand{\bu}{$\star\;$}
\newcommand{\de}{\delta}
\newcommand{\ga}{\gamma}
\newcommand{\ochi}{\overline\chi}
\newcommand{\ba}{\begin{array}}
\newcommand{\ea}{\end{array}}
------
\def\setR{\mathbb{R}}
\def\setN{\mathbb{N}}
\def\setC{\mathbb{C}}
\def\calH {{\cal H}}
\def\ie {{i.e.}}
\def\sgn{\mathrm{sgn}}
\newcommand{\norm}[1]{\parallel\!#1\!\parallel}
\newcommand{\ket}[1]{\mid\!#1\,\rangle}
\newcommand{\bra}[1]{\langle\,#1\!\mid}
\newcommand{\braket}[2]{\langle \, #1 \mid #2 \,\rangle}
\newcommand{\ddroi}[3]{\frac{d^{#1} {#2}}{d{#3}^{#1}}}
\newcommand{\dron}[3]{\frac{\partial^{#1} {#2}}{\partial{#3}^{#1}}}
\newcommand{\e}[1]{e^{#1}}
\newcommand{\sss}[1]{\scriptscriptstyle #1}
\newcommand{\bs}[1]{\boldsymbol{#1}}
\newcommand{\mathcalr}[1]{{\ensuremath{\mathscr{#1}}}}
\newcommand{\dd}{{\rm d}}
\newcommand{\GN}{G_{_\mathrm{N}}}
\newcommand{\lp}{\ell_{_\mathrm{Pl}}}
\newcommand{\fracc}[2]{\frac{\textstyle{#1}}{\textstyle{#2}}}


\title{Mach or Higgs? The mechanisms to generate mass\footnote{To be presented at the XIV Brazilian
School of Cosmology and Gravitation. Rio de Janeiro, 2010}}

\author {\textbf{M. Novello}}
\affiliation{ICRA-Brasil and ICRANet-Italy\footnote{Instituto de
Cosmologia Relatividade e Astrofisica (ICRA/CBPF) - Rio de Janeiro,
Brasil and International Center for Relativistic Astrophysics
(ICRANet)- Pescara, Italy} \\ August 2010
\vspace{1.50cm}\\
Abstract \\
The purpose of this work is to show that the gravitational
interaction is able \\ to generate mass for all bodies. The
condition for this is the existence of an energy distribution
represented by the vacuum or the cosmological constant term $
\Lambda \, g_{\mu\nu}.$ We review briefly the alternative Higgs
mechanism in order to compare both processes.}

\maketitle
\newcommand{\beq}{\begin{equation}}
\newcommand{\eeq}{\end{equation}}
\newcommand{\vare}{\varepsilon}
\newpage
\tableofcontents
\newpage
\section{Introduction}
  The purpose of these notes is to compare the two known mechanisms to generate mass of the elementary constituents
  of all bodies, the basic bricks of which will be taken as representations of the Lorentz-Poincar\'e  group and we
  will analyze them as scalar, spinor, vector and tensor fields. We shall see that in both cases the origin of the mass of any body
  $ \mathbb{A}$ depends on its interaction with its surroundings yielding an overall effect (described either as a scalar field -- in the case
  of the Higgs mechanism -- or as the metric tensor of the geometry of space-time - in the
  case of the gravitational origin) on  $ \mathbb{A}$ which is represented by
  a distribution of energy given by the form
  \begin{equation}
  T_{\mu\nu} = \lambda \, g_{\mu\nu}
\label{29julho}  \end{equation}

  In the literature concerning General Relativity this form of
  energy-momentum tensor is attributed to the cosmological constant
  introduced by Einstein in order to be able to construct a
  model for the geometry of the universe. In the realm of quantum
  field theory, such distribution is identified to the vacuum.
  It is true that if one considers the Machian point of view that the inertia of a
  body $ \mathbb{A} $  depends on the energy distribution of all others bodies in
  the universe, then $ \lambda $ is to be interpreted as the
  cosmological constant \cite{influence}.

These two mechanisms that contemplate the possibility of determining
the mass of any body from elementary principles, are associated to
two distinct universal interactions driven by one of the two fields:
\begin{itemize}
\item{Gravitational field;}
\item{Scalar field.}
\end{itemize}

The idea of using a scalar field to be at the origin of the mass
appeared in the domain of high energy physics and it received the
name "Higgs mechanism". For the time being there is no evidence of
the existence of such scalar in Nature and huge experiments - the
LHC experiment -- are at this very moment in the verge to be
realized in order to prove that such scalar field exists
\cite{veltman}, \cite{vizgin}.

On the other hand, the relationship of mass with gravity is a very
old one and its deep connection has been emphasized in a qualitative
way a huge number of times. We will concentrate our analysis only on
a particular process that admits a systematic realization and allows
for a quantification.

Although the theory of General Relativity may be understood as
completely independent from the Machian idea that inertia of a body
$\mathbb{A}$ is related to the global distribution of energy of all
particles existing in the universe, we must recognize its historical
value in the making the ideology that enabled Einstein to start his
journey toward the construction of a theory of gravitation
 \cite{dicke}.

During the 20th century, the idea of associating the dependence of
local characteristics of matter with the global state of the
universe came up now and then but without producing any reliable
mechanism that could support such proposal \cite{bouncing}. Even the
concept of mass -- that pervades all gravitational processes -- did
not find a realization of such dependence on global structure of the
universe. On the contrary, the most efficient mechanism and one that
has performed an important role in the field of microphysics came
from elsewhere, namely the attempt to unify forces of a
non-gravitational character, such as long-range electrodynamics with
decaying phenomena described by weak interaction. Indeed, the Higgs
model produced an efficient scenario for generating mass to the
vector bosons \cite{halzen} that goes in the opposite direction of
the proposal of Mach. This mechanism starts with the transformation
of a global symmetry into a local one and the corresponding presence
of vector gauge fields. Then, a particular form of the dynamics
represented by $ L_{int}(\varphi)$ of self-interaction of an
associated scalar field in its fundamental state represented by an
energy-momentum tensor given by $ T_{\mu\nu} = L_{int}(\varphi_{0})
\, g_{\mu\nu}$ appears as the vehicle which provides mass to the
gauge fields.

Recently a new mechanism for generation of mass that is a
realization of Mach\rq s idea was proposed \cite{novello}. The
strategy used is to couple the field (scalar, spinor
\cite{novello2}, vector \cite{novello3} and tensor) non minimally to
gravity through the presence of terms involving explicitly the
curvature of space-time. The distribution of the vacuum energy of
the rest-of-the-universe is represented by a cosmological term
$\Lambda.$ The effect of $ \Lambda$ by the intermediary of the
dynamics of the metric of space-time in the realm of General
Relativity is precisely to give mass to the field. Although this
mass depends on the cosmological constant, its value cannot be
obtained a priori \cite{comentario}.

\section{The Higgs proposal}

Consider a theory of a real scalar field $ \varphi$ described by the
Lagrangian
\begin{equation}
\mathbb{L} = \frac{1}{2} \,\partial_{\mu} \varphi \, \partial^{\mu}
\varphi - V(\varphi) \label{741}
\end{equation}
where the potential has the form
$$ V = \frac{1}{2} \, \mu^{2} \, \varphi^{2} + \frac{\lambda}{4} \,
\varphi^{4} $$ In the homogeneous case, in order to satisfy the
equation of motion, the field must be in an extremum of the
potential, which is true for two classes of solution: either
$$ \varphi = 0 $$ or
$$ \varphi_{0}^{2} = - \, \frac{\mu^{2}}{\lambda}. $$
In order to be a minimum the constant $ \mu^{2} $ must be negative.
This is a problem, once it should imply that the mass of the scalar
field is imaginary! However, one can avoid this difficulty in the
following manner. Let us redefine the field by introducing a new
real variable $ \chi $ :
$$ \varphi = \varphi_{0} + \chi,$$
where  $ \varphi_{0} $ is a constant. Substituting this definition
on Lagrangian (\ref{741}) it follows
\begin{equation}
\mathbb{L} = \frac{1}{2} \,\partial_{\mu} \chi \,
\partial^{\mu} \chi + \mu^{2} \, \chi^{2} - \frac{\lambda}{4} \,
\chi^{4} - \lambda \, \varphi_{0} \, \chi^{3}  + \frac{\mu^{4}}{4
\lambda} \label{743}
\end{equation}
This Lagrangian represents a real scalar field $ \chi $ with real
positive mass $ m^{2} = - \, \mu^{2} $ and extra terms of
self-interaction. Note that in the Lagrangian it appears a residual
constant term representing a background constant negative energy
distribution
$$ T_{\mu\nu}(residual) =- \frac{\mu^{4}}{4 \lambda} \, g_{\mu\nu}$$

In the realm of high energy physics it is considered that such term
... " has no physical consequences and can be dropped"
\cite{zumino}. We will come back to this when we analyze its
gravitational effects.

Note that now, the potential of field $ \chi $ takes the form
$$ V = m^{2} \, \chi^{2} + \frac{\lambda}{4} \, \chi^{4} + \lambda \,
\varphi_{0} \, \chi^{3} + constant $$ Its minimum occurs for $ \chi
= 0.$ The others two extrema that exists for constant values $
\chi_{0} $ are points of maxima. The expansion of the field must be
made (for all calculations) around $ \chi = 0$ and not around $
\varphi = 0.$ The reason is that this last is an unstable point and
the series will suffer from convergence. Finally, we note that the
actual field $ \chi $ has a real positive mass.

\subsection{The case of complex field}
Let us now turn to the case of a complex field. The Lagrangian for $
\phi = \phi_{1} + i \, \phi_{2} $ is given by
\begin{equation}
\mathbb{L} = \partial_{\mu} \phi^{*} \, \partial^{\mu} \phi -
V(\phi^{*} \, \phi) \label{741bis}
\end{equation}
where the potential has the form
$$ V =  \mu^{2} \, \phi^{*} \, \phi + \lambda \, (\phi^{*}  \, \phi)^{2} $$

It is convenient to write the field as
$$
\phi = \frac{1}{\sqrt{2}} \, (\phi_{0} + \chi) \,
\exp{\frac{i}{\phi_{0}} \, \theta(x)} $$ The Lagrangian then becomes
\begin{eqnarray}
\mathbb{L} &=& \frac{1}{2} \, \partial_{\mu} \chi \, \partial^{\mu} \chi +
\frac{1}{2} \, \frac{(\phi_{0} + \chi)^{2}}{\phi_{0}^{2}} \, \, \partial_{\mu} \theta \, \partial^{\mu} \theta \nonumber \\
&-& \frac{\mu^{2}}{2} \, (\phi_{0} + \chi)^{2}) - \frac{h}{4} \,
(\phi_{0} + \chi)^{4}
\end{eqnarray}
The extremum of the potential occurs for $ \phi_{0} + \chi = 0.$ For
$ \mu^{2} > 0 $ this extremum is  a minimum.

\subsection{From global to local symmetry}

The theory of the complex field $ \phi $ has a gauge invariance
under the constant map
$$ \phi' = e^{i \alpha} \, \phi.$$
This means that this transformation occurs in everyplace and does
not distinguishes any point of space-time. If the parameter $ \alpha$ becomes
space-time dependent the symmetry is broken. In order to restore the symmetry, one can
use the freedom of the electromagnetic field $ A_{\mu}$ and couple this map with the map
$$ A_{\mu}^{'} = A_{\mu} - \frac{1}{e} \, \partial_{\mu} \alpha. $$ This scheme was
generalized for more general maps (non-abelian theory) by Yang and
Mills in the early 1954 for nonlinear fields, called generically
gauge fields. It is immediate to show  that by minimal coupling of
the scalar field with a gauge field the symmetry is restored. The
modification consists in the passage from a global symmetry (valid
for transformations that are the same  everywhere) to a local
symmetry that depends on the space-time location. A global property
turns into a local one. It is like going from cosmological framework
 -- that deals with the global structure of space-time --- to
microphysics.

\subsection{Mass for a vectorial boson}

The interaction of the complex field $ \phi$ with a vector $ W_{\mu}
$ through the substitution of the derivatives of the scalar field $
\partial_{\mu} \, \phi$ by $ (\partial_{\mu} - i e W_{\mu}) \, \phi
$ using the minimum coupling principle,
 preserves the gauge invariance when the parameter $\alpha $ becomes a
 function of space-time $ \alpha(x).$ This means that the dynamics is invariant under the map

$$ \phi' = \phi \exp{i \, \alpha(x)} $$

$$ W_{\mu}' = W_{\mu} + \frac{1}{e} \, \partial_{\mu} \alpha $$

The Lagrangian, after the above substitution of the field $ \phi =
\phi_{0} + \chi $ turns into
\begin{eqnarray}
\mathbb{L} &=& - \,\frac{1}{4} \, F^{\mu\nu} \, F_{\mu\nu}  +
\frac{1}{2} \, e^{2} \, (\phi_{0} + \chi)^{2}  \, W_{\nu} \, W^{\nu} \nonumber \\
&+& \frac{1}{2} \, \partial_{\mu} \chi \, \partial^{\mu} \chi - V ( \chi) \nonumber \\
&+& \frac{1}{2} \, (\phi_{0} + \chi)^{2} \,
 \partial_{\mu} \theta \, \partial^{\mu} \theta
\end{eqnarray}
Note that this represents the interaction of two real scalar fields
$ \chi $ and $ \theta$ but only the real field $ \chi $ interacts
with the massive vector boson. Due to the gauge invariance, one can
contemplate the possibility  of choosing
$$ \alpha = - \, \frac{\theta}{\phi_{0}} $$ and eliminate $ \theta.$ The dynamics turn into
\begin{eqnarray}
\mathbb{L} &=& - \,\frac{1}{4} \, F^{\mu\nu} \, F_{\mu\nu}  +
\frac{e^{2} \, \phi_{0}^{2}}{2} \, W_{\nu} \, W^{\nu} \nonumber \\
&+& \frac{1}{2} \, \partial_{\mu} \chi \, \partial^{\mu} \chi - V ( \chi) \nonumber \\
&+& ( e^{2} \, \varphi_{0} \, \chi + \frac{e^{2}}{2} \, \chi^{2} ) \, W_{\nu} \, W^{\nu}
\end{eqnarray}
that represents a massive vector field interacting non-minimally
with a real scalar field. Note that one of the degree of freedom of
the theory -- represented by the scalar field $ \theta(x) $ --- was
eliminated. Indeed, it was transformed into an extra degree of
freedom of the massive vector field (that gained one more degree of
liberty going from 2 to 3). The total number we had (two for the
field $ \phi $ and two for the massless field $ W_{\mu})$ is
preserved. It only changed the place. The degree of freedom of $
\theta$ was conceded to the (now) massive vector boson.

It is not difficult to generalize the above procedure for more than
one vector field in such a way that one of them remains massless.
This was the procedure for the case of the unified field theory of
electro-weak interaction: the intermediary boson gain a mass but the
photon remains massless.

\subsection{Mass for a fermion}

Let us couple this scalar field with a spinor $ \Psi$ through the
Lagrangian
\begin{equation}
\mathbb{L} = \frac{1}{2} \,\partial_{\mu} \varphi \, \partial^{\mu}
\varphi - V(\varphi) + \mathbb{L}_{D} - h \, \varphi \, \bar{\Psi}
\, \Psi \label{744}
\end{equation}
where $ \mathbb{L}_{D} $ is Dirac dynamics for massless free field.
Making the same replacement we made previously using $ \chi$ instead
of $ \varphi$ this theory becomes
\begin{equation}
\mathbb{L} = \frac{1}{2} \,\partial_{\mu} \chi \, \partial^{\mu}
\chi - V(\chi) + \mathbb{L}_{D} - h \, (\varphi_{0} + \chi)  \,
\bar{\Psi} \, \Psi \label{745}
\end{equation}
The equation for the spinor field becomes
\begin{equation}
i\gamma^{\mu} \partial_{\mu} \, \Psi - h \, \varphi_{0} \, \Psi - h
\, \chi  \, \bar{\Psi} \, \Psi = 0 \label{746}
\end{equation}
which represents a spinor field of mass $ h \, \varphi_{0} > 0 $
interacting with a scalar field $ \chi.$


\section{Who gives mass to the scalar field that give mass for the
vector and spinor fields? }

In the precedent sections we described the Higgs model that produced
an efficient scenario for generating mass to the vector bosons in
the realm of high-energy physics.  At its origin appears a process
relating the transformation of a global symmetry into a local one
and the corresponding presence of vector gauge fields.

 This mechanism appeals to the intervention of a scalar field
 that appears as the vehicle which provides mass to the  gauge
 vector field  $ W_{\mu}.$   For the mass to be a real and constant value (a different value
for each field) the scalar field $ \varphi$  must be in a minimum
state of its potential $ V .$  This fundamental state of the
self-interacting scalar field has an energy distribution given by $
T_{\mu\nu} = V(\varphi_{0}) \, g_{\mu\nu}.$ A particular form of
self-interaction  of the scalar field $ \varphi$ allows the
existence of a constant value $ V(\varphi_{0}) $ that is directly
related to the mass of $ W_{\mu}.$ This scalar field has its own
mass, the origin of which rests unclear. In \cite{novello} a new
mechanism depending on the gravitational interaction, that can
provides mass to the scalar field was presented. In these lectures
we shall analyze this mechanism.

Although the concept of mass pervades most of all analysis involving
gravitational interaction, it is an uncomfortable situation that
still to this day there has been no successful attempt to derive a
mechanism by means of which mass is understood a direct consequence
of a dynamical process depending on gravity \cite{narlikar}.

The main idea concerning inertia in the realm of gravity according
to the origins of General Relativity, goes in the opposite direction
of the mechanism that we analyzed in the previous section in the
territory of the high-energy physics. Indeed, while the Higgs
mechanism explores the reduction of a global symmetry into a local
one, the Mach principle suggests a cosmical dependence of local
properties, making the origin of the mass of a given body to depend
on the structure of the whole universe. In this way, there ought to
exist a mechanism by means of which this quantity - the mass --
depends on the state of the universe. How to understand such broad
concept of mass? Let us describe an example of such mechanism in
order to see how this vague idea can achieve a qualitative scheme
\cite{lambda}.

\subsection{Mass for scalar field: a trivial case}

We start by considering Mach principle as the statement according to
which the inertial properties of a body $\mathbb{A }$ are determined
by the energy-momentum throughout all space. How could we describe
such universal state that takes into account the whole contribution
of the rest-of-the-universe onto $\mathbb{A }$ ? There is no simpler
way than consider this state as the most homogeneous one and relate
it to what Einstein attributed to the cosmological constant or, in
modern language, the vacuum of all remaining bodies. This means to
describe the energy-momentum distribution of all complementary
bodies of $\mathbb{A }$ as
$$ T_{\mu\nu} = \lambda
\, g_{\mu\nu}
$$

Let $\varphi $ be a massless field the dynamics of which is given by
the Lagrangian $$ L_{\varphi} = \frac{1}{2} \,
\partial_{\alpha} \varphi \, \partial^{\alpha} \varphi$$
In the framework of General Relativity its gravitational interaction
is given by the Lagrangian

\begin{equation}
L = \frac{1}{\kappa_{0}} \, R + \frac{1}{2} \,
\partial_{\alpha} \varphi \, \partial^{\alpha} \varphi + B(\varphi)
\, R - \frac{\lambda}{\kappa_{0}}
\end{equation}
where for the time being the dependence of $ B $ on the scalar field
is not fixed. This dynamics represents a scalar field non-minimally
coupled to gravity. The cosmological constant is added by the
reasons presented above and  represents the influence of the
rest-of-the-universe on $\varphi.$ We shall see that $ \lambda $ is
the real responsible to provide mass for the scalar field. This
means that if we set $ \lambda = 0$ the mass of the scalar field
should vanish.

Independent variation of $\varphi$ and $g_{\mu\nu}$ yields
\begin{equation}
\Box \varphi - R \, B' = 0 \label{3}
\end{equation}

\begin{equation}
\alpha_{0} \, ( R_{\mu\nu} - \frac{1}{2} \, R \, g_{\mu\nu} ) = -
T_{\mu\nu}
 \label{4}
\end{equation}
where we set  $ \alpha_{0} \equiv 2 /\kappa_{0}$ and $ B' \equiv
\partial B / \partial \varphi.$ The energy-momentum tensor is given
by
\begin{eqnarray}
T_{\mu\nu} &=& \partial_{\mu} \varphi \, \partial_{\nu} \varphi -
\frac{1}{2} \, \partial_{\alpha} \varphi \, \partial^{\alpha}
\varphi \, g_{\mu\nu} \nonumber \\
&+& 2 B \, ( R_{\mu\nu} - \frac{1}{2} \, R \, g_{\mu\nu} ) \nonumber
\\
&+& 2 \nabla_{\mu} \nabla_{\nu} B - 2 \Box  B \, g_{\mu\nu} +
\frac{\lambda}{\kappa_{0}}  \, g_{\mu\nu} \end{eqnarray}

Taking the trace of equation (\ref{4}) we obtain
\begin{equation}
( \alpha_{0} + 2 B ) \, R = - \,\partial_{\alpha} \varphi \,
\partial^{\alpha} \varphi - 6 \Box B + \frac{4 \lambda}{\kappa_{0}}
\end{equation}

Inserting this result on the equation (\ref{3} ) yields
\begin{equation}
\Box \varphi + \mathbb{Z} = 0
\end{equation}
where
$$ \mathbb{Z} \equiv \frac{B'}{\alpha_{0} + 2 B} \, \left(
\partial_{\alpha} \varphi \,
\partial^{\alpha} \varphi + 6 \Box B - \frac{4 \lambda}{\kappa_{0}} \right)
$$
or, equivalently,
$$ \mathbb{Z} = \frac{B'}{\alpha_{0} + 2 B} \, \left(
\partial_{\alpha} \varphi \,
\partial^{\alpha} \varphi ( 1 + 6 B'') + 6 \, B'\,\Box \varphi - \frac{4 \lambda}{\kappa_{0}} \right) $$
Therefore, the scalar field acquires an effective self-interaction
through the non-minimal coupling with the gravitational field.  At
this stage it is worth to select among all possible candidates of $
B$ a particular one that makes the factor on the gradient of the
field to disappear in the expression of $ \mathbb{Z} $ by setting
$$ B = a + b \, \varphi - \frac{1}{12} \,\varphi^{2} $$
where $a $ and $ b$ are arbitrary parameters. The quantity $a$
 makes only a re-normalization of the constant $1/ \kappa_{0}$ and
 parameter $b$ is responsible for distinguishing different masses
 for different fields. Making a translation on the field
$$ \Phi  = - \varphi + 6 b $$
it follows
\begin{equation}
\Box \Phi + \mu^{2} \, \Phi = 0 \label{10}
\end{equation}
where \begin{equation} \mu^{2} = \frac{2 \lambda}{3} \,
\frac{\kappa_{ren}}{\kappa_{0}}. \label{5}
\end{equation}
where  $$ \kappa_{ren} = \frac{1}{\alpha_{0} + 2 a + 6 b^{2}}
$$

Thus as a result of the above process the scalar field acquires a
mass $ \mu $ that depends on $ \lambda.$ If $ \lambda $ vanishes
then the mass of the field vanishes. The net effect of the
non-minimal coupling of gravity with the scalar field corresponds to
a specific self-interaction of the scalar field. The mass of the
field appears only if we take into account the existence of all
remaining bodies in the universe in the state in which all existing
matter is on the corresponding vacuum. The values of different
masses for different fields is contemplated in the parameter $b.$

\subsection{Mass for scalar field-II}

Let us now analyze a more general scenario to provide mass to a
scalar field. We start from the Lagrangian that describes a massless
field $\varphi $ that is

 $$ L_{\varphi} = \frac{1}{2} \,
\partial_{\alpha} \varphi \, \partial^{\alpha} \varphi$$
The gravitational interaction yields the modified Lagrangian

\begin{equation}
L = \frac{1}{\kappa} \, R + \frac{1}{2} \, W(\varphi) \,
\partial_{\alpha} \varphi \, \partial^{\alpha} \varphi +  B(\varphi)
\, R - \frac{1}{\kappa} \, \Lambda \label{210}
\end{equation}
where for the time being the dependence of $ B $ and $ W $ on the
scalar field is not fixed. We set $\hbar = c = 1.$

This dynamics represents a scalar field coupled non-minimally with
gravity.  There is no direct interaction between $ \varphi $ and the
rest-of-the-universe (ROTU), except through the intermediary of
gravity described by a cosmological constant $ \Lambda. $ Thus $
\Lambda $ represents the whole influence of the ROTU on $\varphi.$


Independent variation of $\varphi$ and $g_{\mu\nu}$ yields
\begin{equation}
W \, \Box \varphi + \frac{1}{2} \, W' \, \partial_{\alpha} \varphi
\,
\partial^{\alpha} \varphi - B' \, R  = 0 \label{211}
\end{equation}

\begin{equation}
\alpha_{0} \, ( R_{\mu\nu} - \frac{1}{2} \, R \, g_{\mu\nu} ) = -
T_{\mu\nu}
 \label{212}
\end{equation}
where  $ \alpha_{0} \equiv 2 / \kappa $ and $ B' \equiv
\partial B / \partial \varphi.$ The energy-momentum tensor defined
by

 $$T_{\mu\nu} = \frac{2}{\sqrt{- g}} \, \frac{\delta ( \sqrt{-g} \,
 L)}{\delta g^{\mu\nu}} $$

is given by
\begin{eqnarray}
T_{\mu\nu} &=& W \, \partial_{\mu} \varphi \, \partial_{\nu} \varphi
- \frac{1}{2} \, W \, \partial_{\alpha} \varphi \, \partial^{\alpha}
\varphi \, g_{\mu\nu} \nonumber \\
&+& 2 B \, ( R_{\mu\nu} - \frac{1}{2} \, R \, g_{\mu\nu} ) \nonumber
\\
&+& 2 \nabla_{\mu} \nabla_{\nu} B - 2 \Box  B \, g_{\mu\nu} +
\frac{1}{\kappa} \,\Lambda \, g_{\mu\nu} \end{eqnarray}

Taking the trace of equation (\ref{212}) we obtain
\begin{equation}
( \alpha_{0} + 2 B ) \, R = - \, \partial_{\alpha} \varphi \,
\partial^{\alpha} \varphi \, ( W + 6 \, B'') - 6 B' \, \Box \varphi + 4 \,
\frac{\Lambda}{\kappa}
\end{equation}
where we used that  $\Box B = B' \, \Box \varphi + B'' \,
\partial_{\alpha} \varphi \, \partial^{\alpha} \varphi.$

 Inserting this result back on the equation (\ref{211} ) yields
\begin{equation}
\mathbb{M} \, \Box \varphi  \,  + \mathbb{N}  \, \partial_{\alpha}
\varphi \, \partial^{\alpha} \varphi -  \mathbb{Q} = 0 \label{215}
\end{equation}
where
$$ \mathbb{M} \equiv W + \frac{6 (B')^{2}}{\alpha_{0} + 2 B} $$

$$ \mathbb{N} \equiv \frac{1}{2} \, W' + \frac{B' \,( W + 6 B'')}{\alpha_{0} + 2 B} $$

$$ \mathbb{Q} = \frac{4 \, \Lambda \,B'}{\kappa \,(\alpha_{0} + 2 B)} $$

Thus, through the non-minimal coupling with the gravitational field
 the scalar field acquires an effective self-interaction.
 At this point it is worth to select among all possible candidates of
$ B$ and $W $ particular ones that makes the factor on the gradient
of the field to disappear on the equation of motion by setting $
\mathbb{N} = 0. $ This condition will give $ W $ as a function of $
B :$
\begin{equation}
W = \frac{2q - 6 (B')^{2}}{\alpha_{0} + 2 B} \label{216}
\end{equation}
where $ q $ is a constant. Inserting this result into the equation
(\ref{215}) yields
\begin{equation}
\Box \varphi  - \frac{2 \, \Lambda}{q \, \kappa} \, B' = 0.
\end{equation}
At this point one is led to set
$$ B = - \frac{\beta}{4} \, \varphi^{2} $$
to obtain
 \begin{equation}
\Box \varphi + \mu^{2} \, \varphi = 0 \label{218}
\end{equation}
where \begin{equation}
 \mu^{2} \equiv \frac{\beta \, \Lambda}{q \, \kappa} \label{217}
\end{equation}
For the function $ W $ we obtain
$$ W = \frac{2 \, q - 3 \, \beta^{2} \,
\varphi^{2}}{2 \, \alpha_{0} - \beta \varphi^{2}}$$

One should set $ 2 q  = \alpha_{0} $ in order to obtain the standard
dynamics in case $ \beta $ vanishes. Using units were $ \hbar = 1 =
c$ we write
$$ \mathbb{L} = \frac{1}{\kappa}\, R +  \frac{2 \, q - 3 \, \beta^{2}
\, \varphi^{2}}{2 \, (2 \, \alpha_{0} -  \beta \varphi^{2})} \,
\partial_{\alpha} \varphi \, \partial^{\alpha} \varphi - \frac{1}{4}
\, \beta \, \varphi^{2} \, R - \frac{\Lambda}{\kappa} $$

Thus as a result of the gravitational interaction the scalar field
acquires a mass $ \mu $ that depends on the constant $\beta $ and on
the existence of $ \Lambda:$
\begin{equation}
\mu^{2} =  \beta \, \Lambda  \label{219}
\end{equation}

If $ \Lambda $ vanishes then the mass of the field vanishes. The net
effect of the non-minimal coupling of gravity with the scalar field
corresponds to a specific self-interaction of the scalar field. The
mass of the field appears only if we take into account the existence
of all remaining bodies in the universe --- represented by the
cosmological constant --- in the state in which all existing matter
is on the corresponding vacuum. The values of different masses for
different fields is contemplated in the parameter $\beta.$

\subsection{Renormalization of the mass}

The effect of the rest-of-the-universe on a massive scalar field can
be analyzed through the same lines as above. Indeed, let us consider
the case in which there is a potential $ V(\varphi) $
\begin{equation}
L = \frac{1}{\kappa} \, R + \frac{W}{2} \, \partial_{\alpha} \varphi
\, \partial^{\alpha} \varphi + B(\varphi) \, R - V(\varphi)
-\frac{\Lambda}{\kappa} \label{1agosto}
\end{equation}
The equation for the scalar field is given by
\begin{equation}
W \, \Box \varphi + \frac{1}{2} \, W^{'} \,
\partial_{\alpha} \,\varphi \,\partial^{\alpha} \, \varphi - B^{'} \, R + V^{'} = 0
\end{equation}
Use the equation for the metric to obtain the scalar of curvature in
terms of the field and $ \Lambda. $ It then follows that terms in $
\partial_{\alpha} \,\varphi \,\partial^{\alpha} \, \varphi $ are absent if we set
$$ W = \frac{2q - 6 \, (B^{'})^{2}}{\alpha_{0} + 2 B} $$
where $ q $ is a constant. For the case in which $ B = - \beta \,
\varphi^{2} /4$ and for the potential
$$ V = \frac{\mu_{0}}{2} \, \varphi^{2} $$
and choosing $ q = 1/ \kappa $ (in order to obtain the standard
equation of the scalar field in case $ B = 0 )$ yields
\begin{equation}
 \Box \varphi + ( \mu_{0}^{2} + \beta \, \Lambda) \,  \varphi
 + \frac{\beta \, \mu_{0}^{2}}{4} \, \kappa \, \varphi^{3} = 0
\end{equation}
This dynamics is equivalent to the case in which the scalar field
shows an effective potential (in absence of gravity) of the form
$$ V_{eff} = (\mu_{0}^{2} + \beta \, \Lambda) \, \frac{\varphi^{2}}{2}  + \frac{\beta \, \mu_{0}^{2} \, \kappa}{16} \,
\varphi^{4} $$

Thus the net effect of the gravitational interaction for the
dynamics driven by (\ref{1agosto}) is to re-normalize the mass from
the bare value $ \mu_{0} $ to the value

$$\mu^{2} = \mu_{0}^{2} + \beta \, \Lambda.$$
We can then contemplate the possibility that \textbf{all bodies
represented by a scalar field could have the same bare mass and as a
consequence of gravitational interaction acquires a split into
different values characterized by the different values of $ \beta.$
This result is not exclusive of the scalar field but is valid for
any field.}

\section{The case of fermions}

Let us now turn our attention to the case of fermions. The massless
 theory for a spinor field is given by Dirac equation:
\begin{equation} i\gamma^{\mu} \partial_{\mu} \, \Psi  = 0 \label{221}
\end{equation}
This equation is invariant under $\gamma^{5} $ transformation. In
order to have mass for the fermion this symmetry must be broken. Who
is the responsible for this?

\vspace{0.50cm}
\textbf{Gravity breaks the symmetry}
\vspace{0.50cm}

Electrodynamics appears in gauge theory as a mechanism that
preserves a symmetry when one pass from a global transformation to a
local one (space-time dependent map). Nothing similar with gravity.
On the contrary, in the generation of mass through the mechanism
that we are analyzing here, gravity is the responsible to break the
symmetry. In the framework of General Relativity the gravitational
interaction of the fermion is driven by the Lagrangian

\begin{eqnarray}
L &=& \frac{i \, \hbar \, c}{2} \bar{\Psi} \gamma^{\mu} \nabla_{\mu}
\Psi -
\frac{i}{2} \nabla_{\mu} \bar{\Psi} \gamma^{\mu} \Psi \nonumber \\
&+& \frac{1}{\kappa} \, R +  V(\Phi) \, R - \frac{1}{\kappa}
 \, \Lambda \nonumber
\\ &+& L_{CT}
\label{3}
\end{eqnarray}
where the non-minimal coupling of the spinor field with gravity is
contained in the term $ V(\Phi)$ that depends on the scalar
$$ \Phi \equiv \bar{\Psi} \, \Psi$$
which preserves the gauge invariance of the theory under the map $
\Psi \rightarrow \exp(i \, \theta) \, \Psi.$ Note that the
dependence on $ \Phi$ on the dynamics of $ \Psi$ breaks the chiral
invariance of the mass-less fermion, a condition that is necessary
for a mass to appear.

 For the time being the dependence of $ V $ on $ \Phi$ is not fixed. We have added
a counter-term $L_{CT}$ for reasons that will be clear later on.  On
the other hand, the form of the counter-term should be guessed, from
the previous analysis that we made for the scalar case, that is we
set
\begin{equation}
L_{CT} = H(\Phi) \, \partial_{\mu} \Phi \, \partial^{\mu} \Phi
\label{222}\end{equation}

This dynamics represents a massless spinor field coupled
non-minimally with gravity. The cosmological constant represents the
influence of the rest-of-the-universe on $\Psi.$

Independent variation of $\Psi$ and $g_{\mu\nu}$ yields
\begin{equation}
 i\gamma^{\mu} \nabla_{\mu} \, \Psi  +  \left( R \, V'  -
 H' \, \partial_{\mu} \Phi \, \partial^{\mu} \Phi  - 2 H \Box \Phi \right) \, \Psi = 0 \label{223}
\end{equation}

\begin{equation}
\alpha_{0} \, ( R_{\mu\nu} - \frac{1}{2} \, R \, g_{\mu\nu} ) = -
T_{\mu\nu}
 \label{224}
\end{equation}
where  $ V' \equiv \partial V / \partial \Phi.$ The energy-momentum
tensor is given by
\begin{eqnarray}
T_{\mu\nu} &=& \frac{i}{4} \, \bar{\Psi} \gamma_{(\mu} \nabla_{\nu)}
\Psi - \frac{i}{4} \nabla_{(\mu} \bar{\Psi} \gamma_{\nu)} \Psi
\nonumber \\
&+& 2 V ( R_{\mu\nu} - \frac{1}{2} \, R \, g_{\mu\nu} ) + 2
\nabla_{\mu} \nabla_{\nu} V - 2 \Box V g_{\mu\nu} \nonumber \\
&+& 2 H \, \partial_{\mu} \Phi \, \partial_{\nu} \Phi - H \,
\partial_{\lambda} \Phi \, \partial^{\lambda} \Phi \, g_{\mu\nu} +
\frac{\alpha_{0}}{2} \, \Lambda \, g_{\mu\nu} \label{225}
 \end{eqnarray}

Taking the trace of equation (\ref{224}) we obtain after some
algebraic manipulation:
\begin{eqnarray}
( \alpha_{0} + 2 V + V') \, R &=& H'\, \Phi  \,\partial_{\alpha}
\Phi \,
\partial^{\alpha} \Phi \nonumber \\
&+& 2 H \, \Phi \, \Box \Phi - 6 \Box V + 2 \, \alpha_{0} \, \Lambda
\end{eqnarray}

Inserting this result back on the equation (\ref{223}) yields
\begin{equation}
 i\gamma^{\mu} \nabla_{\mu} \, \Psi  +
\left( \mathbb{X} \,  \partial_{\lambda} \Phi \, \partial^{\lambda}
\Phi + \mathbb{Y} \, \Box \Phi \, \right) \, \Psi + \mathbb{Z} \,
\Psi = 0 \label{226}
\end{equation}
where
$$ \mathbb{Z} \equiv \frac{2 \, \alpha_{0} \, \Lambda \, V'}
{Q}
$$

$$ \mathbb{X} = \frac{V'\, ( \Phi \, H' - 2 H - 6 V'')}{Q}
 - H' $$

$$  \mathbb{Y} = \frac{V'\, ( 2 H \, \Phi - 6 V')}{Q}
 - 2 H  $$

where $ Q \equiv \alpha_{0} + 2 V + \Phi \, V'.$

At this stage it is worth selecting among all possible candidates of
$ V$ and $ H $ particular ones that makes the factor on the gradient
and on $ \Box $ of the field to disappear from equation (\ref{226}).
The simplest way is to set  $ \mathbb{X} = \mathbb{Y} = 0$ which
imply only one condition, that is

\begin{equation}
H = \frac{- \, 3 (V')^{2}}{\alpha_{0} + 2 V} \label{227}
\end{equation}

The non-minimal term $ V $ is such that $ \mathbb{Z}$ reduces to a
constant, that is
\begin{equation}
V = \frac{\alpha_{0}}{2} \, \left[ (1 + \sigma \, \Phi)^{-2} - 1
\right] \label{228}
\end{equation}
Then it follows immediately that
\begin{equation}
H = - 3 \alpha_{0} \, \sigma^{2} \,  (1 +  \sigma \, \Phi)^{-4}
\label{229}
\end{equation}
where $ \sigma $ is a constant.

The equation for the spinor becomes
\begin{equation} i\gamma^{\mu} \nabla_{\mu} \, \Psi  - m \Psi= 0 \label{15}
\end{equation}
where \begin{equation}
 m = \frac{4 \, \sigma \, \Lambda}{\kappa \, c^{2}}.
 \label{30julho13}
 \end{equation}

Thus as a result of the above process  the spinor field acquires a
mass $ m $ that depends crucially on the existence of $ \Lambda.$ If
$ \Lambda $ vanishes then the mass of the field vanishes. The
 non-minimal coupling of gravity with the spinor field corresponds to
a specific self-interaction. The mass of the field appears only if
we take into account the existence of all remaining bodies in the
universe --- represented by the cosmological constant. The values of
different masses for different fields are contemplated in the
parameter $ \sigma.$

The various steps of our mechanism can be synthesized as follows:
\begin{itemize}
\item{The dynamics of a massles spinor field $ \Psi$ is described
by the Lagrangian
$$ L_{D} = \frac{i}{2} \bar{\Psi} \gamma^{\mu} \nabla_{\mu} \Psi -
\frac{i}{2} \nabla_{\mu} \bar{\Psi} \gamma^{\mu} \Psi ; $$}
\item{Gravity is described in General Relativity by the scalar of
curvature $$ L_{E} = R ;$$ }
\item{The field interacts with gravity in a non-minimal way described
by the term
$$ L_{int} = V(\Phi) \, R $$
where $ \Phi = \bar{\Psi} \, \Psi ;$ }
\item{The action of the rest-of-the-universe on the spinor field,
through the gravitational intermediary, is contained in the form of
an additional constant term on the Lagrangian noted as $ \Lambda ;$
}
\item{A counter-term depending on the invariant $ \Phi $
is introduced to kill extra terms coming from gravitational
interaction;}
\item{As a result of this process, after specifying $ V $ and $ H $ the field acquires a mass being
described  as
$$ i\gamma^{\mu} \nabla_{\mu} \, \Psi  - m \Psi= 0   $$ where
$ m $ is given by equation (\ref{30julho13}) and is zero only if the
cosmological constant vanishes.}
\end{itemize}

 This procedure allows us to state that the mechanism proposed here is to be
understood as a realization of Mach principle according to which the
inertia of a body depends on the background of the
rest-of-the-universe. This strategy can be applied in a more general
context in support of the idea that (local) properties of
microphysics may depend on the (global) properties of the universe.
We will analyze this in the next session (see also \cite{novello2}).

Thus, collecting all these terms we obtain the final form of the
Lagrangian

\begin{eqnarray}
L &=& \frac{i}{2} \bar{\Psi} \gamma^{\mu} \nabla_{\mu} \Psi -
\frac{i}{2} \nabla_{\mu} \bar{\Psi} \gamma^{\mu} \Psi \nonumber \\
&+& \frac{1}{\kappa} \,  (1 + \sigma \, \Phi)^{-2} \, R -
\frac{1}{\kappa}
 \, \Lambda \nonumber
\\ &-&    \, \frac{6}{\kappa} \, \sigma^{2} \,  (1 +  \sigma \, \Phi)^{-4} \, \partial_{\mu} \Phi \, \partial^{\mu} \Phi
\label{3}
\end{eqnarray}


 \textbf{Some comments}
\begin{itemize}
\item{In the case $ \sigma = 0 $ the Lagrangian reduces to a massless fermion
satisfying Dirac\rq s dynamics plus the gravitational field
described by General Relativity;}
\item{The dimensionality of $ \sigma $ is $L^{3}$; }
\item{The ratio $ m / \sigma = 4 \, \Lambda / \kappa \, c^{2}$ which has the meaning of a density of mass is an universal constant. How to
interpret such universality? }
\end{itemize}


\section{The case of vector fields}
We start with a scenario in which there are only three ingredients:
a massless vector field, the gravitational field and an homogeneous
distribution of energy - that is identified with the vacuum. The
theory is specified by the Lagrangian
\begin{equation}
L = - \frac{1}{4} \, F_{\mu\nu} \, F^{\mu\nu} + \frac{1}{\kappa} \,
R - \frac{\Lambda}{\kappa} \label{13abril1}
\end{equation}
The corresponding equations of motion are

$$F^{\mu\nu}{}{}_{; \nu} = 0 $$

and

$$ \alpha_{0} \, ( R_{\mu\nu} - \frac{1}{2} \, R \, g_{\mu\nu} ) = -
T_{\mu\nu} $$ where $F_{\mu\nu} = \nabla_{\nu} W_{\mu} -
\nabla_{\mu} W_{\nu}$ and  $ \alpha_{0} \equiv 2 / \kappa.$

In this theory, the vacuum $ \Lambda $ is invisible for $ W_{\mu}.$
The energy distribution represented by $ \Lambda$ interacts with the
vector field only indirectly once it modifies the geometry of
space-time. In the Higgs mechanism this vacuum is associated to a
fundamental state of a scalar field $ \varphi$ and it is transformed
in a mass term for $ W_{\mu}.$ The role of $ \Lambda$ is displayed
by the value of the potential $ V(\varphi)$ in its homogeneous
state. We will now show that there is no needs to introduce any
extra scalar field by using the universal character of gravitational
interaction to generate mass for $ W_{\mu}. $

 The point of departure is the recognition that gravity may be
 the real responsible for breaking the gauge symmetry. For this, we modify the
 above Lagrangian to include a non-minimal coupling of the field $
 W_{\mu} $  to gravity in order to explicitly break such
 invariance. There are only two possible ways for this
 \cite{novellobscg}. The total Lagrangian must be of the form

\begin{eqnarray}
\mathbb{L} &=& - \frac{1}{4} \, F_{\mu\nu} \, F^{\mu\nu} + \frac{1}{\kappa} \, R \nonumber \\
 &+&  \frac{\gamma}{6} \,  R \, \Phi  + \gamma \,
R_{\mu\nu} \, W^{\mu} \, W^{\nu} \nonumber \\
 &-&
\frac{\Lambda}{\kappa} \label{210}
\end{eqnarray}
where we define

$$ \Phi \equiv  W_{\mu} \, W^{\mu}. $$

 The first two terms of  $ \mathbb{L} $
represents the free part of the vector and the gravitational fields.
The second line represents the non-minimal coupling interaction of
the vector field with gravity. The parameter $ \sigma $ is
dimensionless. The vacuum -- represented by $\Lambda$ -- is added by
the reasons presented above and it must be understood as the
definition of the expression "the influence of the
rest-of-the-universe on $W_{\mu}".$ We will not make any further
hypothesis on this \cite{14abril}.

 In the present proposed mechanism,  $
\Lambda $ is the real responsible to provide mass for the vector
field. This means that if we set $ \Lambda = 0$ the mass of $
W_{\mu} $ will vanish.

Independent variation of $W_{\mu} $ and $g_{\mu\nu}$ yields
\begin{equation}
F^{\mu\nu}{}{}_{; \nu} + \frac{\gamma}{3} \, R  \, W^{\mu} + 2
\gamma \, R^{\mu\nu} \, W_{\nu} = 0 \label{8abril1}
\end{equation}

\begin{equation}
\alpha_{0} \, ( R_{\mu\nu} - \frac{1}{2} \, R \, g_{\mu\nu} ) = -
T_{\mu\nu}
 \label{8abril2}
\end{equation}
The energy-momentum tensor defined by
 $$T_{\mu\nu} = \frac{2}{\sqrt{- g}} \, \frac{\delta ( \sqrt{-g} \,
 L)}{\delta g^{\mu\nu}} $$
is given by
\begin{eqnarray}
T_{\mu\nu} &=& E_{\mu\nu} \nonumber \\
&+& \frac{\gamma}{3} \,  \nabla_{\mu} \nabla_{\nu} \Phi -
\frac{\gamma}{3} \, \Box  \Phi \, g_{\mu\nu} + \frac{\gamma}{3} \,
\Phi \, ( R_{\mu\nu} - \frac{1}{2} \, R \,
g_{\mu\nu} ) \nonumber \\
&+& \frac{\gamma}{3} \, R W_{\mu} \, W_{\nu}  + 2 \gamma
R_{\mu}{}^{\lambda} \, W_{\lambda} \, W_{\nu} + 2 \gamma
R_{\nu}{}^{\lambda} \, W_{\lambda} \, W_{\mu}  \nonumber
\\ &-& \gamma \, R_{\alpha\beta} \, W^{\alpha} \,W^{\beta} \,
g_{\mu\nu} - \gamma \, \nabla_{\alpha} \, \nabla_{\beta} \, (
W^{\alpha} \, W^{\beta} ) \, g_{\mu\nu} \nonumber \\
&+& \gamma \, \nabla_{\nu} \, \nabla_{\beta} ( W_{\mu} \, W^{\beta})
+ \gamma \, \nabla_{\mu} \, \nabla_{\beta} ( W_{\nu} \, W^{\beta})
\nonumber \\ &+& \gamma \, \Box ( W_{\mu} W_{\nu} ) +
\frac{1}{\kappa} \,\Lambda \, g_{\mu\nu} \end{eqnarray}

where
$$ E_{\mu\nu} =  F_{\mu\alpha} \, F^{\alpha}{}_{\nu} + \frac{1}{4}
F_{\alpha\beta} \, F^{\alpha\beta} \, g_{\mu\nu}
$$

Taking the trace of equation (\ref{8abril2}) we obtain
\begin{equation}
R =  2 \, \Lambda - \kappa \, \gamma \, \nabla_{\alpha} \,
\nabla_{\beta} \, ( W^{\alpha} \, W^{\beta} )
\end{equation}
Then, using this result back into equation (\ref{8abril1}) it
follows
\begin{eqnarray}
F^{\mu\nu}{}{}_{; \nu} &+&  \frac{2\, \gamma \, \Lambda}{3} \,
W^{\mu}
 \nonumber \\
 &-& \frac{\kappa \, \gamma^{2}}{3} \,  \nabla_{\alpha} \,
\nabla_{\beta} \, ( W^{\alpha} \, W^{\beta}) \, W^{\mu} \nonumber \\
&+& 2 \, \gamma \, R^{\mu}{}_{\nu} \, W^{\nu} = 0
\end{eqnarray}
 The non-minimal coupling with gravity yields an effective
self-interaction of the vector field and a term that represents its
direct interaction with the curvature of space-time. Besides, as a
result of this process the vector field acquires a mass $ \mu $ that
depends on the constant $\gamma $ and on the existence of $
\lambda.$ The term

$$ 2 \, \gamma \, R^{\mu}{}_{\nu} \, W^{\nu} $$
gives a contribution (through the dynamics of the metric equation
(\ref{8abril2}) of $ \gamma \, \Lambda$ yielding for the mass the
formula
\begin{equation} \mu^{2} =  \frac{5}{3} \, \gamma \, \Lambda
\label{219}
\end{equation}
Note that the Newton\rq s constant does not appear in our formula
for the mass.  The net effect of the non-minimal coupling of gravity
with $ W^{\mu} $ corresponds to a specific self-interaction of the
vector field. The mass of the field appears only if we take into
account the existence of the rest-of-the-universe --- represented by
$ \Lambda$
--- in the state in which this environment is on the
corresponding vacuum. If $ \Lambda $ vanishes then the mass of the
field vanishes.The values of different masses for different fields
are contemplated in the parameter $\gamma.$

\vspace{0.50cm}
 \textbf{Quantum perturbations}
 \vspace{0.50cm}

How this process that we have been examining here to give mass to
all kind of bodies should be modified in a quantum version? We note,
first of all, that the gravitational field is to be treated at a
classical level, once there is neither theoretical nor observational
evidence that exists a quantum version of gravitational interaction.
Thus, any modification of the present scheme means to introduce
quantum aspects of the vector field. This will not change the whole
scheme of generation of mass described above.Indeed, in the
semi-classical approach in which the matter field is quantized but
the metric is not, the modification of the equation of general
relativity becomes
\begin{equation}
\alpha_{0} \, ( R_{\mu\nu} - \frac{1}{2} \, R \, g_{\mu\nu} ) = -
 \, <T_{\mu\nu}>
 \label{4}
\end{equation}
where the field is in a given specific state. Throughout all the
process of gravitational interaction the system does not change its
state, allowing the same classical treatment as above.


\section{The case of spin-two field}

As in the previous cases we start with a scenario in which there are
only three ingredients: a linear tensor field, the gravitational
field and an homogeneous distribution of energy identified with the
vacuum. We note that there are two possible equivalent ways to
describe a spin-two field that is:

\begin{itemize}
  \item {Einstein frame}
  \item {Fierz frame}
\end{itemize}
 according we use a symmetric second order tensor $ \varphi_{\mu\nu}$ or
 the third-order tensor tensor $ F_{\alpha\beta\lambda}.$
 Although the Fierz representation is not used for most of the
works dealing with spin-2 field, it is far better than the Einstein
frame when dealing in a curved space-time\cite{antunes}. Thus, let
us review briefly the basic properties of the Fierz
frame\footnote{We use the notation $A_{(\alpha}B_{\beta )} =
A_\alpha B_\beta + A_\beta B_\alpha$, $A_{[\alpha}B_{\beta ]} =
A_\alpha B_\beta - A_\beta B_\alpha$.}. We start by defining a
three-index tensor $F_{\alpha\beta\mu}$ which is anti-symmetric in
the first pair of indices and obeys the cyclic identity:
\begin{equation}
F_{\alpha\mu\nu} + F_{\mu\alpha\nu} = 0, \label{01}
\end{equation}
\begin{equation}
F_{\alpha\mu\nu} + F_{\mu\nu\alpha} + F_{\nu\alpha\mu} = 0.
\label{02}
\end{equation}
This expression implies that the dual of $F_{\alpha\mu\nu}$ is
trace-free:
\begin{equation}
\stackrel{*}{F}{}^{\alpha\mu}{}_{\mu} = 0 , \label{02bis}
\end{equation}
where the asterisk represents the dual operator, defined in terms of
$\eta_{\alpha\beta\mu\nu}$ by
\[
\stackrel{*}{F}{}^{\alpha\mu}{}_{\lambda} \equiv \frac{1}{2} \,
\eta^{\alpha\mu}{}_{\nu\sigma}\,F^{\nu\sigma}{}_{\lambda}.
\]
The tensor $F_{\alpha\mu\nu}$ has 20 independent components. The
necessary and sufficient condition for $F_{\alpha\mu\nu}$ to
represent an unique spin-2 field (described by 10 components) is
\footnote{Note that this condition is analogous to that necessary
for the existence of a potential $A_{\mu}$ for the EM field, given
by $\stackrel{*}{A}{}^{\alpha\mu}{}_{,\alpha} = 0.$}
\begin{equation}
\stackrel{*}{F}{}^{\alpha (\mu\nu)}{}_{,\alpha} = 0, \label{03}
\end{equation}
which can be rewritten as
\begin{eqnarray}
&& {{F_{\alpha\beta}}^{\lambda}}{}_{,\mu} + {{F_{\beta\mu}}^{\lambda}}%
{}_{,\alpha} + {{F_{\mu\alpha}}^{\lambda}}{}_{,\beta} -\frac{1}{2}
\delta^{\lambda}_{\alpha} (F_{\mu ,\beta} - F_{\beta ,\mu}) + \nonumber \\
&& - \frac{1}{2} \delta^{\lambda}_{\mu} (F_{\beta ,\alpha} -
F_{\alpha ,\beta}) - \frac{1}{2} \delta^{\lambda}_{\beta} (F_{\alpha
,\mu} - F_{\mu ,\alpha}) = 0.
\end{eqnarray}
A direct consequence of the above equation is the identity:
\begin{equation}
F^{\alpha\beta\mu}{}_{\, ,\mu} = 0 \ .  \label{z1}
\end{equation}
We call a tensor that satisfies the conditions given in the
Eqns.(\ref{01}), (\ref{02}) and (\ref{03}) a Fierz tensor. If
$F_{\alpha\mu\nu}$ is a Fierz tensor, it represents an unique spin-2
field. Condition (\ref{03}) yields a connection between the Einstein
frame (EF) and the Fierz frame (FF): it implies that there exists a
symmetric second-order tensor $\varphi_{\mu\nu}$ that acts as a
potential for the field. We write
\begin{eqnarray}
2\,F_{\alpha\mu\nu} &=& \varphi_{\nu [\alpha,\mu ]} + \left(
\varphi_{,\alpha} - \varphi_{\alpha}{}^{\lambda}{}_{,\lambda}
\right)\, \eta_{\mu\nu}\nonumber \\
 &-& \left(\varphi_{,\mu} -
\varphi_{\mu}{}^{\lambda}{}_{,\lambda} \right)\, \eta_{\alpha\nu} .
\label{04}
\end{eqnarray}
where $\eta_{\mu\nu}$ is the flat spacetime metric tensor, and the
factor $2$ in the l.h.s. is introduced for convenience.

Taking the trace of equation (\ref{04}) $F_{\alpha}\equiv
F_{\alpha\mu\nu}\eta^{\mu\nu}$ it follows that
$$
F_{\alpha} = \varphi_{,\alpha} -
\varphi_{\alpha}{}^{\lambda}{}_{,\lambda},
$$
where . Thus we can write
\begin{equation}
2 F_{\alpha\mu\nu} = \varphi_{\nu [\alpha,\mu ]}  + F_{[\alpha }
\,\eta_{\mu ]\nu}.  \label{06}
\end{equation}

Using the properties of the Fierz tensor we obtain the important
identity:
\begin{equation}
F^{\alpha }{}_{(\mu \nu ),\alpha }\equiv - \, 2 \, G^{(L)}{}_{\mu
\nu } , \label{07}
\end{equation}
where $G^{(L)}{}_{\mu \nu }$ is the linearized Einstein tensor,
defined by the perturbation $ g_{\mu\nu} = \eta_{\mu\nu} + \varphi
_{\mu \nu }$ by
\begin{equation}
2 \, G^{(L)}{}_{\mu \nu }\equiv \Box \,\varphi _{\mu \nu }-\varphi
^{\epsilon }{}_{(\mu ,\nu )\,,\epsilon }+\varphi _{,\mu \nu }-\eta
_{\mu \nu }\,\left( \Box \varphi -\varphi ^{\alpha \beta
}{}_{,\alpha \beta }\right) . \label{08}
\end{equation}

The divergence of $F^{\alpha }{}_{(\mu \nu ),\alpha }$ yields Bianci
 identity:

\begin{equation}
F^{\alpha (\mu \nu )}{}_{,\alpha \mu }\equiv 0.  \label{07bis}
\end{equation}
Indeed,
\begin{equation}
F^{\alpha \mu \nu }{}_{,\alpha \mu }+F^{\alpha \nu \mu }{}_{,\mu
\alpha }=0. \label{070}
\end{equation}
The first term vanishes identically due to the symmetric properties
of the field and the second term vanishes due to equation
(\ref{z1}). Using Eqn.(\ref{07}) the identity which states that the
linearized Einstein tensor $G^{(L)}{}_{\mu \nu }$ is divergence-free
is recovered.

We shall build now dynamical equations for the free Fierz tensor in
flat spacetime. Our considerations will be restricted here to linear
dynamics. The most general theory can be constructed from a
combination of the three invariants involving the field. These are
represented by $A$, $B$ and $W$:
$$
A \equiv F_{\alpha \mu \nu }\hspace{0.5mm}F^{\alpha \mu \nu } ,
\;\;\;\;\;\;\;\;B \equiv F_{\mu }\hspace{0.5mm}F^{\mu },
$$
$$
W \equiv F_{\alpha \beta \lambda }\stackrel{\ast }{F}{}^{\alpha
\beta \lambda }=\frac{1}{2}\,F_{\alpha \beta \lambda
}\hspace{0.5mm}F^{\mu \nu \lambda }\,\eta ^{\alpha \beta }{}_{\mu
\nu } .
$$
$W$ is a topological invariant so we shall use only the invariants
$A$ and $B$. The EOM  for the massless spin-2 field in the ER is
given by

\begin{equation}
G^{(L)}{}_{\mu \nu }=0.  \label{014bis}
\end{equation}
As we have seen above, in terms of the field $F^{\lambda \mu \nu}$
this equation can be written as

\begin{equation}
F^{\lambda (\mu \nu )}{}_{,\lambda }=0.  \label{014}
\end{equation}
The corresponding action takes the form

\begin{equation}
S=\frac{1}{k}\,\int {\rm d}^{4}x\,(A-B) . \label{013}
\end{equation}
  Then,
\begin{equation}
\delta S=\int F^{\alpha\, (\mu \nu) }{}_{,\alpha }\,\delta \varphi
_{\mu \nu}\,d^{4}x . \label{018}
\end{equation}
 we obtain
\begin{equation}
\delta S=-2 \int G^{(L)}{}_{\mu \nu
}\,\delta\varphi^{\mu\nu}\,d^{4}x, \label{018bis}
\end{equation}
where $G^{(L)}\mbox{}_{\mu \nu }$ is given in Eqn.(\ref{08}).

Let us consider now the massive case. If we include a mass for the
spin 2 field in the Fierz frame, the Lagrangian takes the form
\begin{equation}
{\cal L}=A-B+\frac{m^{2}}{2}\,\left(\varphi _{\mu \nu }\, \varphi
^{\mu \nu}-\varphi ^{2}\right),  \label{30julho1356}
\end{equation}
and the EOM that follow are
\begin{equation}
F^{\alpha }{}_{(\mu \nu ),\alpha } - m^{2}\,\left(
\varphi_{\mu\nu}-\varphi\,\eta _{\mu \nu }\right) =0 , \label{mc1}
\end{equation}
or equivalently,
\[
G^{(L)}{}_{\mu \nu } +\frac{m^{2}}{2} \,\left( \varphi _{\mu \nu
}-\varphi \,\eta _{\mu\nu }\right) =0.
\]
The trace of this equation gives
\begin{equation}
F^{\alpha }{}_{,\alpha }+ \frac{3}{2}\,m^{2}\,\varphi =0,
  \label{mc12}
\end{equation}
while the divergence of Eqn.(\ref{mc1}) yields
\begin{equation}
F_{\mu }=0.  \label{mc121}
\end{equation}
This result together with the trace equation gives $\varphi =0.$

In terms of the potential, Eqn.(\ref{mc121}) is equivalent to
\begin{equation}
\varphi _{,\,\mu }-\varphi ^{\epsilon }{}_{\mu \,,\epsilon }=0.
\label{mcd121}
\end{equation}
It follows that we must have
\[
\varphi ^{\mu \nu }{}_{,\nu }=0.
\]
Thus we have shown that the original ten degrees of freedom (DOF) of
$F_{\alpha\beta\mu}$ have been reduced to five (which is the correct
number for a massive spin-2 field) by means of the  five constraints
\beq \varphi ^{\mu \nu }{}_{,\nu }=0,\;\;\;\;\;\;\;\;\;\varphi = 0.
\label{fc} \eeq

\vspace{0.50cm}
 \textbf{Equation of spin-2 in curved background}
\vspace{0.50cm}

 The passage of the spin-2 field equation   from Minkowski spacetime to
arbitrary curved riemannian manifold presents ambiguities due to the
presence of second order derivatives of the rank two symmetric
tensor $\varphi_{\mu\nu}$ that is used in the so called
Einstein-frame (see for instance \cite{Aragone-Deser-2}). These
ambiguities disappear when we pass to the Fierz frame representation
that deals with the three index tensor $F_{\alpha\mu\nu}$ as it was
 shown in \cite{Novello-Neves}.

 There results  a unique form of minimal
coupling, free of ambiguities. Let us define from $\varphi_{\mu\nu}$
two auxiliary fields $G^{(I)}{}_{\mu \nu}$ and $G^{(II)}{}_{\mu \nu
}$ through the expressions:
\begin{align}
\nonumber 2 \, G^{(I)}{}_{\mu \nu }&\equiv Ê\\Ê \Box \,\varphi _{\mu
\nu }-\varphi_{\epsilon(\mu ;\nu )}{}^{;\epsilon }&+\varphi _{;\mu
\nu }-\eta _{\mu \nu }\,\left( \Box \varphi -\varphi ^{\alpha \beta
}{}_{;\alpha \beta }\right) , \label{28nov1}\\
\nonumber 2 \, G^{(II)}{}_{\mu \nu }&\equiv  \\
\Box \,\varphi _{\mu \nu }-\varphi_{\epsilon(\mu}{}^{;\epsilon}{}
_{;\nu)}&+\varphi _{;\mu \nu }-\eta _{\mu \nu }\,\left( \Box \varphi
-\varphi ^{\alpha \beta }{}_{;\alpha \beta }\right). \label{28nov2}
\end{align}
These objects differ only in the order of the second derivative in
the second term on the r.h.s. of the above equations. The equation
of motion \cite{30julho1356} free of ambiguities concerns the tensor
field
\begin{equation}
\widehat{G}_{\mu \nu } \equiv \frac{1}{2} \, \left( G^{(I)}{}_{\mu
\nu } + G^{(II)}{}_{\mu \nu } \right) \label{28nov3}
\end{equation}
and is given by
\begin{equation}
\widehat{G}_{\mu \nu } + \frac{1}{2} \,m^{2} \, (\varphi_{\mu \nu}
-\varphi g_{\mu\nu})= 0. \label{28nov4}
\end{equation}
which is precisely the usual equations for massive spin-2 field.
\vspace{0.50cm}

\textbf{Generating mass for the spin-2 field} \vspace{0.50cm}

We follow the same strategy as in the previous case and take the
dynamics of the spin-2 field as given by

\begin{eqnarray}
\mathbb{L} &=&  F_{\alpha\mu\nu} \, F^{\alpha\mu\nu} - F_{\alpha} \, F^{\alpha} + \frac{1}{\kappa} \, R \nonumber \\
 &+&  a \, R_{\alpha\mu\beta\nu} \, \varphi^{\alpha\beta} \, \varphi^{\mu\nu}  \nonumber \\
 &-& \frac{\Lambda}{\kappa} \label{21julho}
\end{eqnarray}

The equations of motion are given by:
\begin{equation}
F^{\alpha }{}_{(\mu \nu ); \alpha }+ 2a \, R_{\alpha\mu\beta\nu} \,
\varphi^{\alpha\beta} =0,
  \label{21julho16}
\end{equation}

\begin{equation}
\frac{1}{\kappa} \, \left( R_{\mu\nu} - \frac{1}{2} \, R \,
g_{\mu\nu} + \frac{\Lambda}{2}\, g_{\mu\nu} \right) + T_{\mu\nu} + a
\, Y_{\mu\nu} = 0 \label{22julho710}
\end{equation}
where the quantity $  Y_{\mu\nu} $ is given by the variation of the
 non minimal coupling term:

\begin{equation} \delta
\int \sqrt{- g} \, R_{\alpha\mu\beta\nu} \, \varphi^{\alpha\beta} \,
\varphi^{\mu\nu} = \, \int \sqrt{- g} \, Y_{\mu\nu} \, \delta
g_{\mu\nu}
 \label{22julho715}\end{equation}
where  $ Y_{\mu\nu} $ is given in terms of $ S_{\alpha\mu\beta\nu} $
defined as
$$ S_{\alpha\mu\beta\nu}  \equiv \varphi_{\alpha\beta} \,
\varphi_{\mu\nu} -\varphi_{\alpha\nu} \, \varphi_{\beta\mu} $$ which
has the symmetries:
$$ S_{\alpha\mu\beta\nu} = - \, S_{\alpha\mu\nu\beta} = - \,
S_{\mu\alpha\beta\nu} = S_{\beta\nu\alpha\mu}.$$ A direct
calculation yields
$$ Y^{\mu\nu} \equiv  S^{\lambda\mu\nu\epsilon}{}{}_{; \epsilon ;
\lambda} - \frac{1}{2} \, R_{\alpha\sigma\beta\lambda} \,
\varphi^{\alpha\beta} \, \varphi^{\sigma\lambda} \,  g^{\mu\nu}  +
\frac{3}{2} \, R_{\alpha\sigma\beta}{}^{( \mu} \,
 \varphi^{\nu) \sigma} \, \varphi^{\alpha\beta}  - \, \frac{1}{2} \, R_{\alpha\sigma\beta\lambda} \,
\varphi^{\alpha\beta} \, \varphi^{\sigma\lambda} \,  g_{\mu\nu}$$

Let us remind that the Riemann curvature can be written in terms of
its irreducible quantities involving the Weyl conformal tensor $
W_{\alpha\sigma\beta\lambda}$ and the contracted Ricci tensor by the
formula:

$$ R_{\alpha\mu\beta\nu} = W_{\alpha\mu\beta\nu} + \frac{1}{2} \left( R_{\alpha\beta} \, g_{\mu\nu}
 + R_{\mu\nu} \, g_{\alpha\beta} - R_{\alpha\nu} \, g_{\beta\mu} - R_{\beta\mu} \,
 g_{\alpha\nu} \right) - \frac{1}{6} \, R \,
 g_{\alpha\mu\beta\nu}.$$
Then
$$ R_{\alpha\mu\beta\nu} \, \varphi^{\alpha\beta} \, \varphi^{\mu\nu}  = W_{\alpha\mu\beta\nu} \,  \varphi^{\alpha\beta} \, \varphi^{\mu\nu}
+  \left( R_{\alpha\beta} - \frac{1}{6} \, R  \, g_{\alpha\beta}
\right) \, \left( \varphi \, \varphi^{\alpha\beta} -
\varphi^{\alpha}{}_{\lambda} \, \varphi^{\lambda\beta} \right).$$ We
can then re-write the equation of the spin-2 field as
\begin{equation}
F^{\alpha }{}_{(\mu \nu ); \alpha } - \frac{a \, \Lambda}{3} \,
(\varphi_{\mu\nu} - \varphi \, g_{\mu\nu} ) + 2a \,
W_{\alpha\mu\beta\nu} \, \varphi^{\alpha\beta} + Q_{\mu\nu} =0,
  \label{27julho18}
\end{equation}
where $ Q_{\mu\nu} $ contain non-linear terms of interaction of the
spin-2 field with gravity.

\section{Generalized Mach\rq s principle}

In this section we present an extension of Mach principle in similar
lines as it has been suggested by Dirac, Hoyle and others. This
generalization aims to produce a mechanism that transforms the vague
idea according to which local properties may depend on the
universe\rq s global characteristics into an efficient process. We
will apply the strategy that we used in the precedent sections to
generate mass in order to elaborate such generalization.

\vspace{0.50cm}

 \textbf{ The cosmological
influence on the microphysical world: the case of chiral-invariant
Heisenberg-Nambu-Jona-Lasinio dynamics}

\vspace{0.50cm}

 There have been many discussions
in the scientific literature in the last decades related to the
cosmic dependence of the fundamental interactions. The most popular
one was the suggestion of Dirac -- the so called Large Number
Hypothesis -- that was converted by Dicke and Brans into a new
theory of gravitation, named the scalar-tensor theory. We will do
not analyze any of these here. On the contrary, we will concentrate
on a specific self-interaction of an elementary field and show that
its correspondent dynamics is a consequence of a dynamical
cosmological process. That is, to show that dynamics of elementary
fields in the realm of microphysics, may depend on the global
structure of the universe.

 The first question we have to face concerns the
choice of the elementary process. There is no better way than start
our analysis with the fundamental theory proposed by Nambu and
Jona-Lasinio concerning a dynamical model of elementary particles
\cite{nambu}. Since the original paper until to-day hundreds of
papers devoted to the NJL model were published \cite{volkov}. For
our purpose here it is enough to analyze the nonlinear equation of
motion that they used in their original paper as the basis of their
theory which is given by

$$ i\gamma^{\mu} \nabla_{\mu} \, \Psi  - 2s ( A + i \, B \, \gamma^{5} )\Psi =
0 $$ This equation, as remarked by these authors, was proposed
earlier by Heisenberg \cite{heisenberg} although in a quite
different context. We will not enter in the analysis of the theory
that follows from this dynamics. Our question here is just this: is
it possible to produce a model such that HNJL
(Heisenberg-Nambu-Jona-Lasinio) equation for
 spinor field becomes a consequence of the gravitational interaction of a free massless Dirac
 field with the rest-of-the-universe? We shall see that the answer is
 yes.

We used Mach\rq s principle as the statement according to which the
inertial properties of a body $\mathbb{A }$ are determined by the
energy-momentum throughout all space. We follow here a similar
procedure and will understand the Extended Mach Principle as the
idea which states that the influence of the rest-of-the-universe on
microphysics can be described through the action of the
energy-momentum distribution identified with the cosmic form
$$
T^{U}_{\mu\nu} = \Lambda \, g_{\mu\nu}
$$

\vspace{0.50cm}
 \textbf{Non minimal coupling with gravity}
\vspace{0.50cm}

 In the framework of General Relativity we set the
dynamics of a fermion field $\Psi$ coupled non-minimally with
gravity to be given by the Lagrangian (we are using units were
$\hbar = c = 1)$

\begin{equation}
L = L_{D} + \frac{1}{\kappa} \, R +  V(X) \, R - \frac{1}{\kappa}
 \, \Lambda + L_{CT}
\label{401}
\end{equation}
where
\begin{equation}
L_{D} \equiv \frac{i}{2} \bar{\Psi} \gamma^{\mu} \nabla_{\mu} \Psi -
\frac{i}{2} \nabla_{\mu} \bar{\Psi} \gamma^{\mu} \Psi
 \label{302}
\end{equation}
The non-minimal coupling of the spinor field with gravity is
contained in the term $ V(X)$ and depends on the scalar $ X$ defined
by  $$ X = A^{2} + B^{2}$$ where $ A = \bar{\Psi} \, \Psi $ and $B =
i \bar{\Psi} \, \gamma^{5} \, \Psi.$ We note that we can write, in
an equivalent way,  $$X = J_{\mu} \, J^{\mu} $$ where $ J^{\mu} =
\bar{\Psi} \gamma^{\mu}  \Psi.$ This quantity $ X $ is chiral
invariant, once it is invariant under the map
$$ \Psi' = \gamma^{5} \, \Psi.$$ Indeed, from this $\gamma^{5}$ transformation,
it follows

$$ A' = - \, A, \,\,   B' = - \, B; \,\,  then,  X' = X.$$

The case in which the theory breaks chiral invariance and the
interacting term  $ V $ depends only on the invariant $A$ -- is the
road to the appearance of a mass as we saw in the previous sections
\cite{novello2}. Here we start from the beginning with a chiral
invariant theory. For the time being the dependence of $ V $ on $ X
$ is not fixed. We have added $L_{CT}$ to counter-balance the terms
of the form $
\partial_{\lambda} X \, \partial^{\lambda} X $ and $ \Box X $
that appear due to the gravitational interaction. The most general
form of this counter-term is
\begin{equation}L_{CT} = H(X) \, \partial_{\mu} X \, \partial^{\mu} X
\label{403}\end{equation} We shall see that $ H $ depends on $ V$
and if we set $ V = 0 $ then $ H $ vanishes. This dynamics
represents a massless spinor field coupled non-minimally with
gravity. The cosmological constant represents the influence of the
rest-of-the-universe on $\Psi.$

Independent variation of $\Psi$ and $g_{\mu\nu}$ yields
\begin{equation}
 i\gamma^{\mu} \nabla_{\mu} \, \Psi  +  \Omega   \, ( A + i \, B \, \gamma^{5} )\Psi = 0 \label{404}
\end{equation}
where
$$  \Omega \equiv 2 R \, V'  -
 2 H' \, \partial_{\mu} X \, \partial^{\mu} X  - 4 H \Box X $$

\begin{equation}
\alpha_{0} \, ( R_{\mu\nu} - \frac{1}{2} \, R \, g_{\mu\nu} ) = -
T_{\mu\nu}
 \label{405}
\end{equation}
where we set  $ \alpha_{0} \equiv 2 /\kappa $ and $ V' \equiv
\partial V / \partial X.$ The energy-momentum tensor is given by
\begin{eqnarray}
T_{\mu\nu} &=& \frac{i}{4} \, \bar{\Psi} \gamma_{(\mu} \nabla_{\nu)}
\Psi - \frac{i}{4} \nabla_{(\mu} \bar{\Psi} \gamma_{\mu)} \Psi
\nonumber \\
&+& 2 V ( R_{\mu\nu} - \frac{1}{2} \, R \, g_{\mu\nu} ) + 2
\nabla_{\mu} \nabla_{\nu} V - 2 \Box V g_{\mu\nu} \nonumber \\
&+& 2 H \, \partial_{\mu} X \, \partial_{\nu} X - H \,
\partial_{\lambda} X \, \partial^{\lambda} X \, g_{\mu\nu} +
\frac{\alpha_{0}}{2} \, \Lambda \, g_{\mu\nu} \label{406}
 \end{eqnarray}

Taking the trace of equation (\ref{405}), after some simplification
and using
\begin{equation}
 \Box V = V' \, \Box X + V'' \, \partial_{\mu} X \, \partial^{\mu}
X
\end{equation}
 it follows
\begin{eqnarray}
( \alpha_{0} + 2 V + 2 \, V' \, X) \, R &=& (4 H X - 6 V') \Box X
\nonumber
\\ &+& ( 2 H'\, X - 6 V'' - 2 H) \, \partial_{\alpha} X \,
\partial^{\alpha} X \nonumber \\
&+& 2 \, \alpha_{0} \, \Lambda
\end{eqnarray}
Then

\begin{eqnarray}
\Omega &=&  \left( \mathbb{M} \, \Box X + \mathbb{N} \,
\partial_{\mu} X \,
\partial^{\mu} X \right) \nonumber
\\
&+& \frac{4 \, \alpha_{0} \, \Lambda \, V'}{\alpha_{0} + 2 V + 2 \,
V' \, X}
\end{eqnarray}
where
$$ \mathbb{M} = \frac{2 V' ( 4 H X - 6 V')}{ \alpha_{0} + 2 V + 2 \, V' \,
X }  - 4 \, H $$

$$ \mathbb{N} = \frac{2 V' \, ( 2 \, X \, H' -  6 V'' - 2 \, H)}{\alpha_{0} + 2 V + 2 \, V' \, X}  - 2 \, H'
$$

Defining $ \Delta \equiv \alpha_{0} + 2 V + 2 \, V' \, X $ we
re-write $\mathbb{M}$ and $\mathbb{N}$ as
$$  \mathbb{M} = - \, \frac{4}{\Delta} \, \left( 3 \, V'^{2} + H \,(\alpha_{0} + 2 V ) \right) $$
$$  \mathbb{N} = - \, \frac{2}{\Delta} \, \left( 3 \, V'^{2} + H \,(\alpha_{0} + 2 V ) \right)' $$

Inserting this result on the equation (\ref{404}) yields
\begin{equation}
 i\gamma^{\mu} \nabla_{\mu} \, \Psi  +
\left( \mathbb{M} \, \Box X  + \mathbb{N} \,
\partial_{\lambda} X \, \partial^{\lambda} X \right) \,
\Psi + \mathbb{Z} \, ( A + i \, B \, \gamma^{5} )\Psi = 0
\label{407}
\end{equation}
where
$$  \mathbb{Z} = \frac{ 4 \,\alpha_{0} \, \Lambda \, V'}{\Delta }
  $$

At this stage it is worth to select among all possible candidates of
$ V$ and $ H $ particular ones that makes the factor on the gradient
and on $ \Box $ of the field to disappear from equation (\ref{407}).

The simplest way is to set  $ \mathbb{M} = \mathbb{N} = 0,$  which
is satisfied if
$$H = - \, \frac{3 \, V'^{2}}{\alpha_{0} + 2 V} $$

Imposing that  $ \mathbb{Z}$ must reduce to a constant we obtain
\begin{equation}
V = \frac{1}{\kappa} \, \left[ \frac{1}{ 1 + \beta \, X} - 1
\right]. \label{408}
\end{equation}
As a consequence of this,
\begin{equation}
H = - \, \frac{3 \, \beta^{2}}{2 \kappa}  \, \frac{1}{(1 +  \beta \,
X)^{3}} \label{409}
\end{equation}
where $ \beta $ is a constant. Using equations \ref{407}) and
\ref{408}) the equation for the spinor becomes

\begin{equation}
 i\gamma^{\mu} \nabla_{\mu} \, \Psi  - 2s ( A + i \, B \, \gamma^{5} )\Psi = 0  \label{410}
\end{equation}
where \begin{equation}
 s = \frac{2 \, \beta \, \Lambda}{\kappa (\hbar \, c)}.
 \label{20}
 \end{equation}

Thus as a result of the gravitational interaction the spinor field
satisfies Heisenberg-Nambu-Jona-Lasinio equation of motion. This is
possible due to the influence of the rest-of-the-Universe on $
\Psi.$  If $ \Lambda $ vanishes then the constant of the
self-interaction of $ \Psi$ vanishes.

The final form of the Lagrangian is provided by
\begin{equation}
 L = L_{D} +
\frac{1}{\kappa \, ( 1 + \beta X )} \, R  - \frac{1}{\kappa}  \,
\Lambda - \frac{ 3 \beta^{2}}{2 \kappa} \, \frac{1}{( 1 + \beta
X)^{3}} \, \partial_{\mu} X \, \partial^{\mu} X \label{15maio}
\end{equation}

In this section we analyzed the influence of all the material
content of the universe on a fermionic field when this content is in
two possible states: in one case its energy distribution is zero; in
another case it is in a vacuum state represented by the homogeneous
distribution $ T_{\mu\nu} = \Lambda g_{\mu\nu}.$ Note that when
$\Lambda $ vanishes,  the dynamics of the field is independent of
the global properties of the universe and it reduces to the massless
Dirac equation

$$ i\gamma^{\mu} \nabla_{\mu} \, \Psi  = 0 $$

In the second case, the rest-of-the-universe induces on field
$\Psi$ the Heisenberg-Nambu-Jona-Lasinio non-linear dynamics

$$  i\gamma^{\mu} \nabla_{\mu} \, \Psi  -
2s \, (A + iB \gamma^{5}) \, \Psi= 0.$$

Such scenario shows a mechanism by means of which the rules of the
microphysical world depends on the global structure of the universe.
It is not hard to envisage others situations in which the above
mechanism can be further applied.

\section{Appendix: Vacuum state in non-linear theories}

Although the cosmological constant was postulated from first
principles, quantum field theory gave a simple interpretation of $
\Lambda $ by its association to the fundamental vacuum state. It is
possible to describe its origin even classically as a consequence of
certain special states of matter. For instance, non linear theories
produce classically a vacuum, defined by its distribution of
energy-momentum tensor provided by expression (\ref{29julho}). Let
us review very briefly how this occurs in a specific example. We
start by the standard definition of the symmetric energy-momentum
tensor as variation of the Lagrangian induced by variation of the
metric tensor, that is

\begin{equation}
T_{\mu\nu} = \frac{2}{\sqrt {-g}} \frac{\delta L \sqrt{- g}}{\delta
g ^{\mu\nu}} \label{10julho740}
\end{equation}

In order to present a specific example, let us concentrate on the
case of electromagnetic field in which the Lagrangian depends only
on the invariant $ F $ defined by
$$ F \equiv F_{\mu\nu} \,  F^{\mu\nu}$$
Then, the expression of the energy-momentum tensor is given by

\begin{equation}
T_{\mu\nu} = -4 \, L_{F} \, F_{\mu}{}^{\alpha} \, F_{\alpha\nu} - L
\, g_{\mu\nu}. \label{NL3}\end{equation} where $ L_{F} = \partial L
/ \partial F$ represents the derivative of the Lagrangian with
respect to the invariant $ F. $ The corresponding equation of motion
of the field is provided by
\begin{equation}
\left( L_{F} \, F^{\mu\nu} \right)_{; \, \nu} = 0.
\end{equation}
where the symbol $ ; $ represents covariant derivative. This
equation admits a particular solution when $ L_{F} $ vanishes for
non-null constant value $ F_{0} $ . When the system is in this
state, the corresponding expression of the energy-momentum tensor
reduces to
$$ T_{\mu\nu} = \Lambda \, g_{\mu\nu} $$
where
$$ \Lambda =  L_{0}.$$

The consequences of this state in Cosmology due to non linear
theories of Electrodynamics was revisited recently (see
 \cite{novello-erico}). A by-product is the emergence of effective
 geometries that mimics gravitational processes like, for
 instance, non-gravitational black holes or analogue expanding universes in  laboratory.

\section*{Acknowledgements}
The major part of this work was done when visiting the International
Center for Relativistic Astrophysics (ICRANet) at Pescara (Italy). I
would extend this acknowledgement to its efficient administration. I
would also thank my colleagues of ICRA-Brasil and particularly Dr
J.M. Salim with whom I exchanged many discussions. This work was
partially supported by {\em Conselho Nacional de Desenvolvimento
Cient\'{\i}fico e Tecnol\'ogico} (CNPq), FINEP and {\em
Funda\c{c}\~ao de Amparo \`a Pesquisa do Estado de Rio de Janeiro}
(FAPERJ) of Brazil.


\begin{thebibliography}{100}
\bibitem{comentario}
Let us remark that it is possible to understand Mach principle in a
broad sense. Indeed, for the method of obtaining mass using the
gravity mechanism, the notion of Totality (or rest-of-the-universe)
 admits two alternative interpretations. This is related to the fact that
 when we deal with the vacuum represented by the distribution of energy by
$ T_{\mu\nu} = \lambda \, g_{\mu\nu}$
 it is completely irrelevant -- for the gravitational
 mechanism of providing mass --- if parameter $\lambda$ has a classical global origin (the Universe)
identified with the cosmological constant introduced by Einstein; or
a local quantum one (the environment) identified with the vacuum of
quantum fields.
\bibitem{lambda}
At this point one should note that the expression
rest-of-the-universe may not have a strict cosmological meaning but
instead is a short term to represent the whole background of matter
that really affects $\mathbb{A}. $
\bibitem{influence}  However this is not mandatory. The term
rest-of-the-universe concerns the environment of $ \mathbb{A}, $
that is the whole domain of influence on $ \mathbb{A} $ of the
remaining bodies in the universe.
\bibitem{14abril} There is not any compelling reason to identify this constant with the actual
cosmological constant or the value of the critical density $ 10^{-
48} Gev^{4}$ provided by cosmology.
\end{thebibliography}
 \end{document}